# Evaporation of water and urea solution in a magnetic field; the role of nuclear isomers

Sruthy Poulose, M. Venkatesan, Matthias Möbius and J.M.D. Coey*

*School of Physics, Trinity College, Dublin 2, Ireland*

*\* jcoey@tcd.ie*

**Abstract.**

*Hypothesis.* Ortho and para water are the two nuclear isomers where the hydrogen protons align to give a total nuclear spin of 0 or 1. The equilibrium ratio of 3:1 is established slowly in freshly evaporated water vapour and the isomers then behave distinct gasses, with their own partial pressures. Magnetic-field-induced ortho↔para transformations are expected to alter the evaporation rate.

*Experiment.* Evaporation from beakers of deionized water and a 6 M solution of urea is monitored simultaneously for periods from 1 to 60 hours with and without a 500 mT magnetic field, while logging the ambient temperature and humidity. The balances with the two beakers are shielded in the same Perspex container. Many runs have been conducted over a two-year period.

*Findings.* The evaporation rate of water is found to increase by 12 ± 7 % of in the field but that of water in urea solution decreases by 28 ± 6 %. Two effects are at play. One is dephasing of the Larmor precession of adjacent protons on a water molecule in a field gradient, which tends to equalize the isomer populations. The other is Lorentz stress on the moving charge dipole, which can increase the proportion of the ortho isomer. From analysis of the time and field dependence of the evaporation, we infer an ortho fraction of 39 ± 1% in fresh vapour from water and 60 ± 5 % in fresh vapour from urea.

**Key words** Magnetic water treatment, Magnetic fields, Halbach magnet, Evaporation of water, Evaporation of urea solution, Ortho and para water molecules, Larmor precession, Lorentz stress, Hydrogen bonding,

## 1. Introduction

Modification of the physical and chemical properties of water by exposure to a magnetic field has been a topic of dispute for years. Most controversial, perhaps, have been claims that by passing hard water once through a non-uniform magnetic field, it is possible to influence the subsequent precipitation of limescale when the water is heated. Such claims were widely dismissed despite a body of experimental evidence to the contrary, because no mechanism could be envisaged whereby such fleeting exposure of hard water to a magnetic field could influence the precipitation of calcium carbonate hours later. The picture changed when it emerged that a significant fraction of the calcium in water was bound in nanoscale polymeric clusters of amorphous soft matter [1,2] that act as seeds for the subsequent nucleation and precipitation of calcium carbonate from supersaturated solution [3]. The new theory of nucleation is under debate [4], but it opens a possibility of understanding long-term consequences of exposing of hard water to a magnetic field in terms of the nuclear spin structure of two protons at the surface of a prenucleation cluster [5].

Here we are concerned with a simpler question. Does the evaporation rate of water change in the presence of a static magnetic field, and if so why? A review of magnetic water treatment by Chibowski and Szcses [6] mentions persistent magnetic field effects that enhance the water evaporation rate [7] [8-12], which is the focus of our paper. There are three types of experiments, First are those where the weight of an open container of water is measured as it is exposed to a static magnetic field $B$ and compared to a no-field reference [8,12-14]. These usually involve interrupting the experiment after different lapses of time to weigh the evaporating water, which may either be left at ambient temperature or heated in an oven [15]. Alternatively, the loss of weight by evaporation can be measured in zero field at different times after a single exposure to a static field [16]. The field is usually produced by a permanent magnet, and it is not uniform in magnitude or direction over the surface of the evaporating water, which is contained in a beaker.

In the second type of experiment, the water is exposed to a nonuniform field, by continuous dynamic flow at a velocity $\gtrsim 1\,\text{ms}^{-1}$ around a circuit where a permanent magnet surrounds a section of the pipe [13,17-19]. The experimental setup resembles that used for magnetic treatment of hard water to control limescale, except that the water is circulated repeatedly through the inhomogeneous magnetic field, not just once. This type of treatment is equivalent to periodically exposing the water to a ~ 20 ms magnetic field pulse at frequency of about 1 Hz. After the circulation period, the magnetically-treated water is removed and its rate of evaporation is tracked by weighing after different intervals of time. Remarkably, both types of experiments give qualitatively similar results. They have been combined in a two-stage experiment to increase the effect[20]. Exposure of water to a magnetic field of a few tens or hundreds of millitesla has been found to significantly increase the subsequent rate of evaporation. This implies that the water retains a memory of its dynamic or static [16] magnetic treatment that persists for times of order an hour or more [12,20]. Various authors use different experimental protocols, and the results depend on temperature, humidity, magnetic field and time in different ways, so there is much inconsistency in the results. The reported increases in evaporation rate associated with the magnetic field range from a few percent to more that 30% [8,11,12,15,16,19,21]. The magnetic field is thought to modify somehow the network of molecular hydrogen bonding in water. There is no agreed explanation of how this happens, but it should be noted that the direct decrease in energy of $½\chi_{mol}B^2/\mu_0$ of water with molar susceptibility [22] $\chi_{mol} = -1.6 \cdot 10^{-10}$, under a field of 1 T = - 64 µJ/mol, is nine orders of magnitude less than 23.3 kJ/mol, the energy of hydrogen bonds in water [23]. It has been suggested [9][24] that the effects of magnetic water treatment are associated with the proportions of ortho and para isomers of $H_2O$, which have net nuclear spins $I = 1$ and $I = 0$, respectively, and an influence of the field on their librational oscillations [24]

The third type of experiment is different. Here the water is exposed to an intense magnetic field in a superconducting solenoid, with a large horizontal field gradient $\nabla_x B$. The product $B\nabla_x B$ is 320 $T^2 m^{-1}$. The explanation here [7] is related to the paramagnetic susceptibility of atmospheric oxygen, which is displaced by the evaporating water vapour. The changes are 17 % of the buoyancy forces on air, which leads to a field-induced modification of the convection and evaporation rate. Similar experiments have been conducted in a vertical field gradient[10]. Buoyancy effects are again observed, and evaporation was found to be faster in microgravity.

In this work we investigate the evaporation rate in a much smaller, quasi-uniform field. We have built an automated setup to monitor the evaporation of water or an aqueous solution continuously for periods of up to 60 h, comparing the evaporation rate in the magnetic field with that of a no-field control, while recording the temperature and humidity. We compare the effect on pure water with that on concentrated solutions of urea, where hydrogen bonding is disrupted by the solute. The results are discussed in terms of possible effects of a magnetic field on the two isomers of water, which behave as quasi-independent gasses in the vapour phase.

## 2. Experimental methods

The experimental uses two identical KERN Model EW-2200 Precision balances (2.2 kg capacity, 10 mg sensitivity) to monitor the weight of water or aqueous solution as they evaporate from two 100 mL beakers. One beaker is exposed to a magnetic field and the other is a no-field control. The balances are placed side by side in a closed polymethyl methacrylate (Perspex) box, as shown in Fig. 1. The mass of water is automatically recorded for periods of 16 or 60 h (over a day or a weekend). No transfer of liquid involved, only a real time record of the change of mass with and without a magnetic field. A substantially-uniform horizontal magnetic field parallel to the liquid surface is produced by a large 8-segment Nd-Fe-B Halbach ring magnet with inner and outer diameters of 130 mm and 330 mm, and height of 95 mm. The in-plane field profile across the bore and the profile along the axis are shown in Supplementary Information. The beaker was centred in the magnet bore and the liquid surface was near the middle of the magnet, The average field over the surface was close to 500 mT, with a weak radial gradient, reaching 3 $Tm^{-1}$ around the edge. The beakers are raised on 140 mm supports above the patens of the balances and the magnet is mounted on a four-legged aluminium stand in a position where its magnetic field has no influence on the balance. Four suspended thermocouples are placed inside and outside each beaker, and there is a temperature/humidity sensor within the box. Readings from both balances and the temperature and humidity sensors collected with a TESTO 174H data logger and recorded in real time using a LabVIEW code. Although there are fluctuations in ambient temperature and relative humidity over the course of a run, the conditions are always similar for both the in-field liquid and the no-field control. The initial quantities of water were 10, 20, 30 and 50 mL and for 6M urea solutions where the urea is non-volatile the quantity used was 68 mL, corresponding to 50 mL of water. Experiments for each quantity of liquid were repeated 4 – 8 times on different days, including experiments where the balances or the position of the magnet were interchanged. Data were collected with the box in different positions in the laboratory. Table 1 summarizes properties of the liquids used.

Table 1. Details of the liquids investigated

| Liquid | | Specific Gravity | Surface tension (mNm$^{-1}$) | pH | Conductivity (Sm$^{-1}$) | Boiling temperature (°C) |
|---|---|---|---|---|---|---|
| Water Millipore | 18.2 MΩcm | 1.000 | 72.0 | 7.6 | < 0.01 | 99.9 |
| Urea solution 99.9% | 6.0 M | 1.095 | 74.4 | 7.9 | < 0.01 | 103.7 |

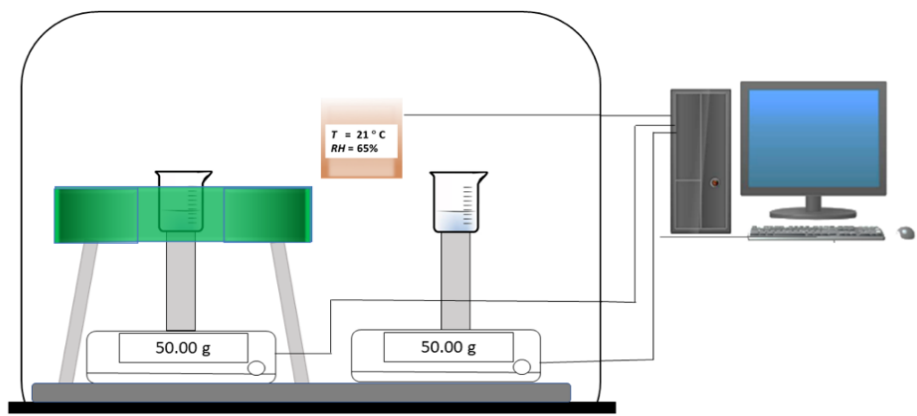

Figure 1: Experimental evaporation setup with in-field and no-field beakers of water, whose masses are measured simultaneously. Magnetic field profiles are in Supplementary Information

## 3. Results

Besides magnetic field, the subject of our investigation, the evaporation rates of water and aqueous solutions depend on temperature, humidity and airflow. Convection above the liquid surface is limited by the walls of the beaker. Temperature and humidity are not controlled, but they are monitored in the Perspex box throughout the duration of each run, and the temperature of the water in each beaker is also monitored throughout. Figure 2 compares results of some runs for different volumes of water, 10, 20 and 50 mL, with mass 10, 20 and 50 g, comparing the no-field reference (black lines) to the water in the magnet (red lines). The average magnetic field over the surface is similar in the three cases and the magnetic field gradient at the surface is less than 3 Tm$^{-1}$. In the course of these experiments, the evaporation rate fluctuated by 10 % or more because of slow variations of ~ 1 °C in ambient temperature and ~ 5% in relative humidity. The data in figure 2d) show that the percentage weight loss after 16 hours varies as the inverse of the mass *m* of a sample, and the rate of evaporation of water is enhanced by 8 ± 2 % in the magnetic field. The proportionality of mass loss to inverse sample mass is expected because evaporation is a surface effect, proportional to the surface area exposed to the magnetic field. In every one of 36 independent runs of 16 or 60 hours, the in-field evaporation rate of water was found to be greater. The evaporation rate averaged over all runs for which *RH* = 0.73 ± 0.06 is 0.0297 ± 0.0081 kgm$^{-2}$h$^{-1}$ without field and 0.0331 ± 0.0088 kgm$^{-2}$h$^{-1}$ in the 500 mT field. Overall, the evaporation rate increases by 12 % in the field with a standard deviation of 7%.

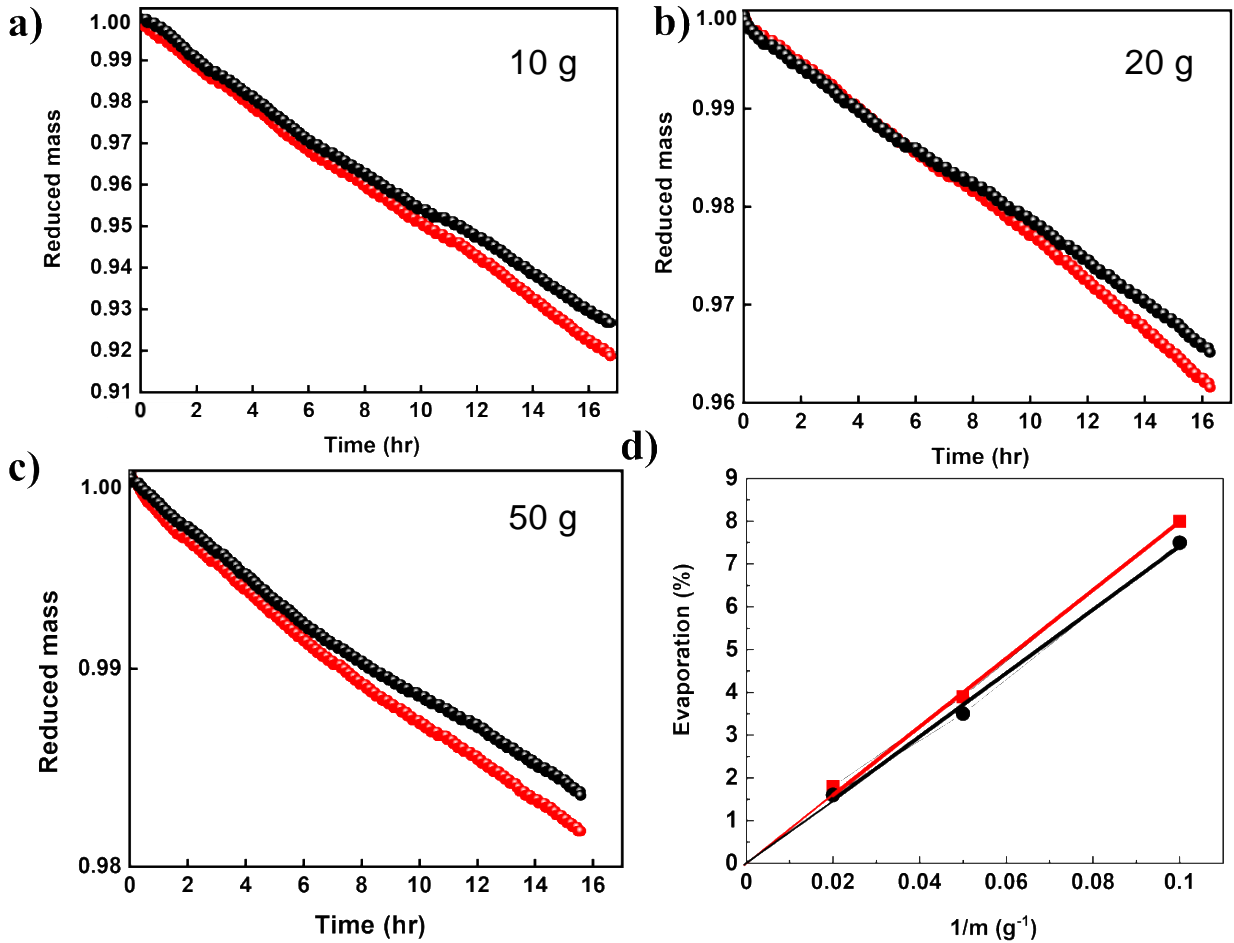

Figure 2: Relative weight loss by evaporation of water in a beaker in a 500 mT magnetic field parallel to the surface (red lines) and a reference with no magnetic field (black lines). The initial masses

are a) 10 g, b) 20 g and c) 50 g. The fourth panel plots the weight loss as a function of inverse sample mass showing that the effect scales with surface area.

In figure 3, we show data for 30 mL of water and 68 mL of 6M Urea in runs where the evaporation rate remained almost constant over the course of the 16 hour experiments, due to smaller fluctuations in ambient temperature and relative humidity, or correlated and opposite drifts in these variables. In Figure 3a) for 30 mL water, the increase in evaporation rate in the field is 19%. The effects on the water and urea solutions in Figs 3a) and 3c) are quite different. In urea, evaporation is inhibited by the magnetic field, being reduced by 23%, whereas in water, it is enhanced. The variations of temperature and humidity are illustrated in Fig 3b) and 3d). Since the vapour pressure of urea is negligible at room temperature and urea forms clusters in water, the weight loss is due to urea-modified water, where the hydrogen bonding within and between clusters of water molecules differs from that in pure water[25].

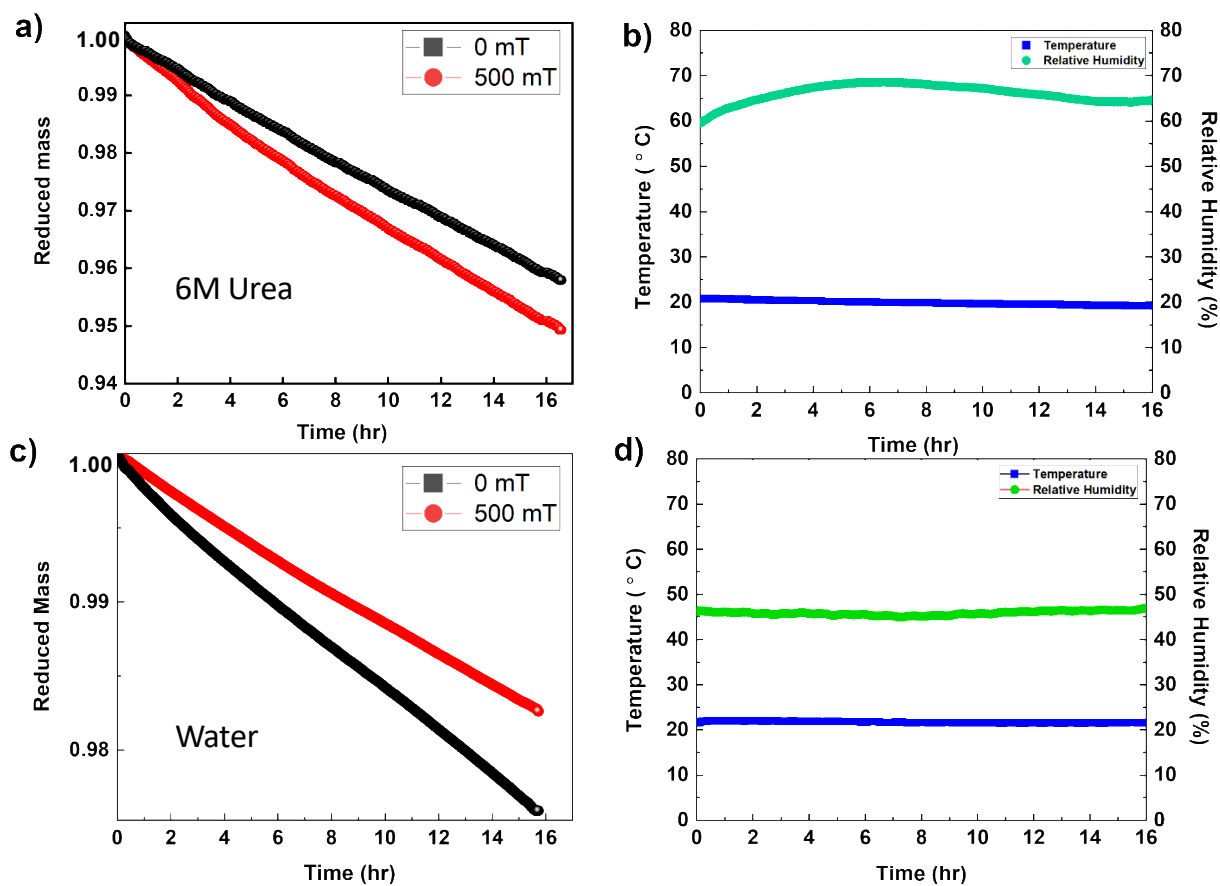

Figure 3: Relative weight loss by evaporation of water in runs where there were fluctuations of temperature and humidity over the 16 h runs were minor a) and b) 30 mL water, c) and d) 68 mL 6 M Urea. Black lines are for reference samples, red lines for samples exposed to the 500 mT magnetic field.

We also analysed more extended runs where the evaporation of 50 mL of liquid was monitored over a period of 60 hours. The ambient temperature and humidity fluctuated during the run, as illustrated in Fig. 4b), but the evaporation rate of the sample in the magnetic field at any time was almost always greater than that of the reference. At the end of this particular experiment with water, the net weight loss in the magnetic field was 15% greater than that of the reference.

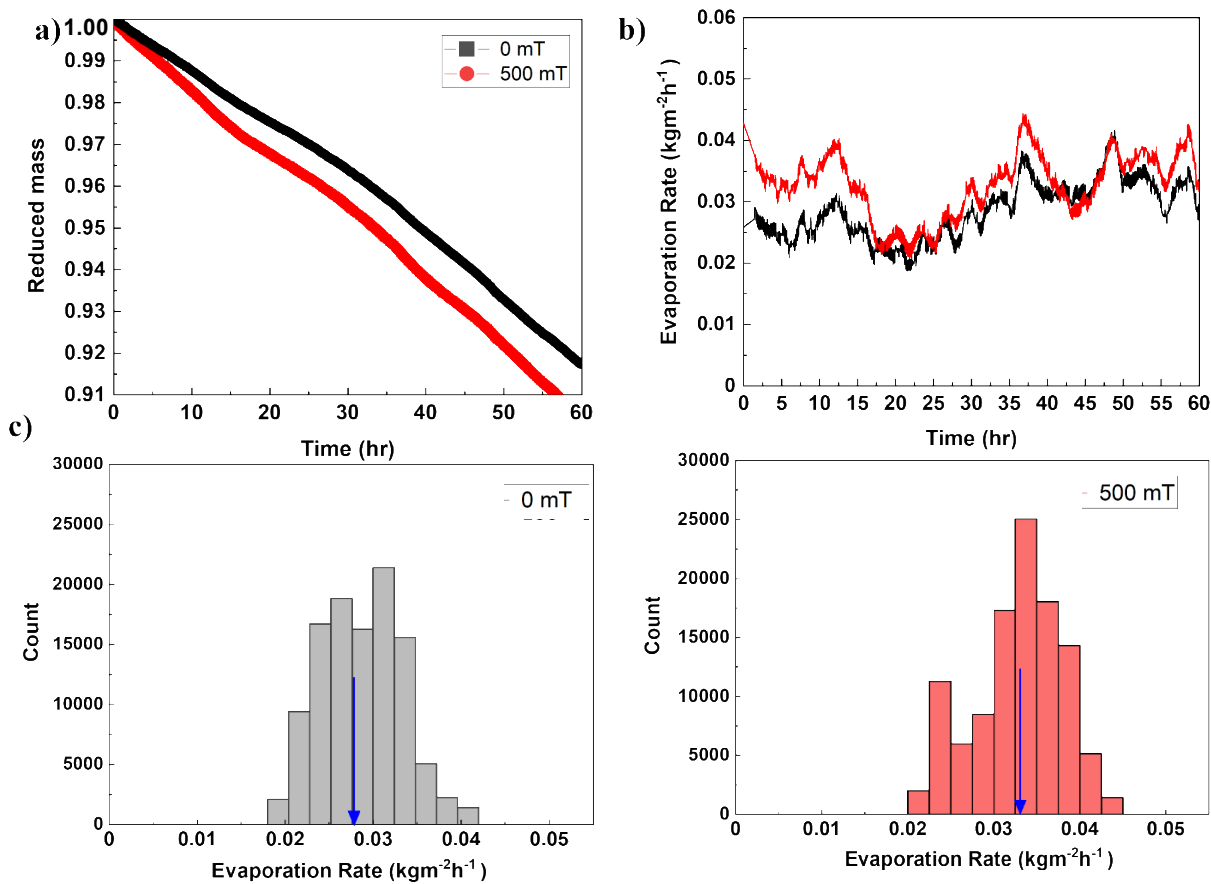

Figure 4: Extended 60 h run of evaporation of water versus time for a 50 mL sample of water in 500 mT (red) and the no-field reference (black, grey). a) relative weight loss by evaporation b) evaporation rates versus time and c) histograms of the evaporation rates, where the average values are marked by blue arrows.

The temperature of the evaporating water was found to increase by about 1°C in the course of a 16 hour run, and the temperature change for the samples in the magnetic field was slightly (0.1 ± 0.1°C) higher, as shown in Supplementary Information, but this does not explain the difference in evaporation rate.

Shorter-term two-hour runs were carried out to follow the initial minutes of evaporation after pouring the deionized water or urea solution into a beaker in a closed balance. Typical results are illustrated in Figure 5. Evaporation in zero field is fastest in the first few minutes for water and it then settles down to less than half the initial rate after about half an hour. The urea solution is quite different. The initial evaporation is very slow, and it becomes three times faster after the initial transient. The data were fitted to the function

$$m(t) = m_0 e^{-t/t_o} + m_1(1 - g_1 t). \tag{1}$$

The mass of water at the start of an experiment is $m(0)$ and $m(t)/m(0)$ is the reduced mass. The time constant of the initial exponential transient, when the water vapour in the air in contact with the water surface has the equilibrium 3:1 ortho/para ratio, is $t_o = 9$ min. At longer times the isomeric ratio will approach that of the freshy-evaporated water. The average of eight one-hour runs for water gave an initial evaporation rate of $0.134 \pm 0.030$ and an average $t_o$ of $14 \pm 3$ min. The long-time evaporation rate is $0.068 \pm 0.010$ kgm$^{-2}$h$^{-1}$. The average of four one-hour runs for urea gave an initial evaporation rate that was over ten times slower than that of water, while the difference in long-time evaporation rate was much smaller, although evaporation of water from urea was still slower than pure water in conditions of comparable humidity (Figure 6 and Table 3). The decay time of the initial transient was $t_0 = 62$ min, much longer than for pure water, so the 'steady' decay times from the one-hour runs are underestimates.

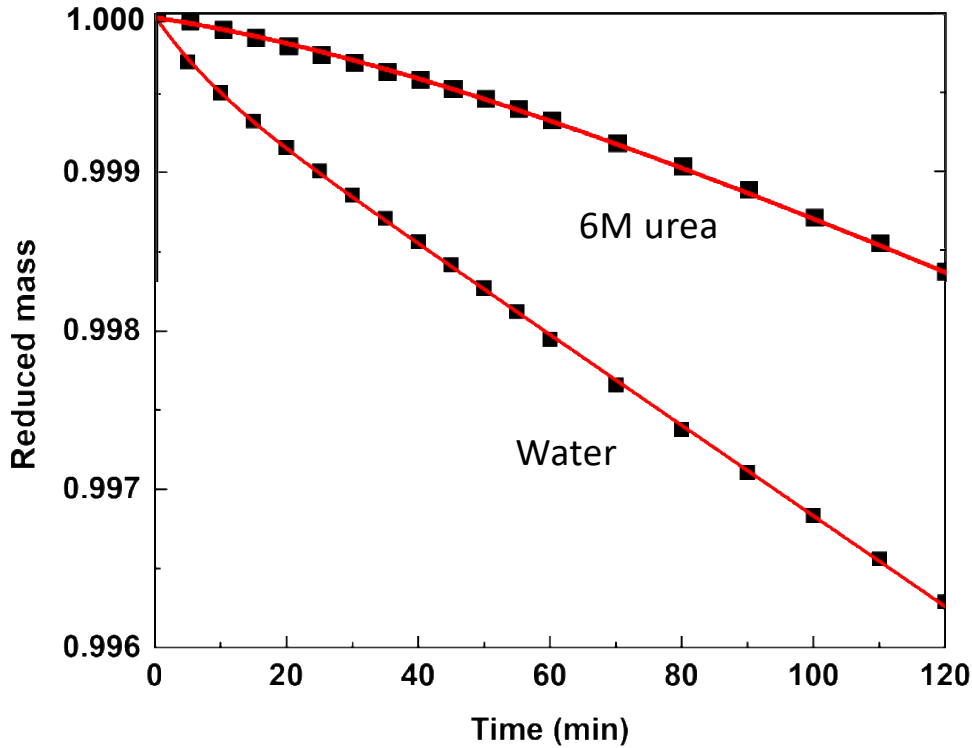

Figure 5. Short-term evaporation of Millipore water and Urea solution in stagnant conditions, showing the steady-state and initial transient regimes. The transient, which decays exponentially with a time constant of 9 min for water and 62 min for urea has opposite signs in the two cases.

The evaporation rate is very sensitive to the ambient relative humidity. Although we did not control it, the value was mostly in the range 60% – 70%. On some days it was quite steady over a 16-hour period, but on others it changed by 5% - 20% in the course of a run, with a corresponding change in evaporation rate. We make use of those data to plot the of change of evaporation rate $\Delta g$ versus the change in relative humidity $\Delta RH$ in Figure 6. The variation of $g$ with $RH$ is nonlinear. The fitted slope in Fig. 6 is 2.2.

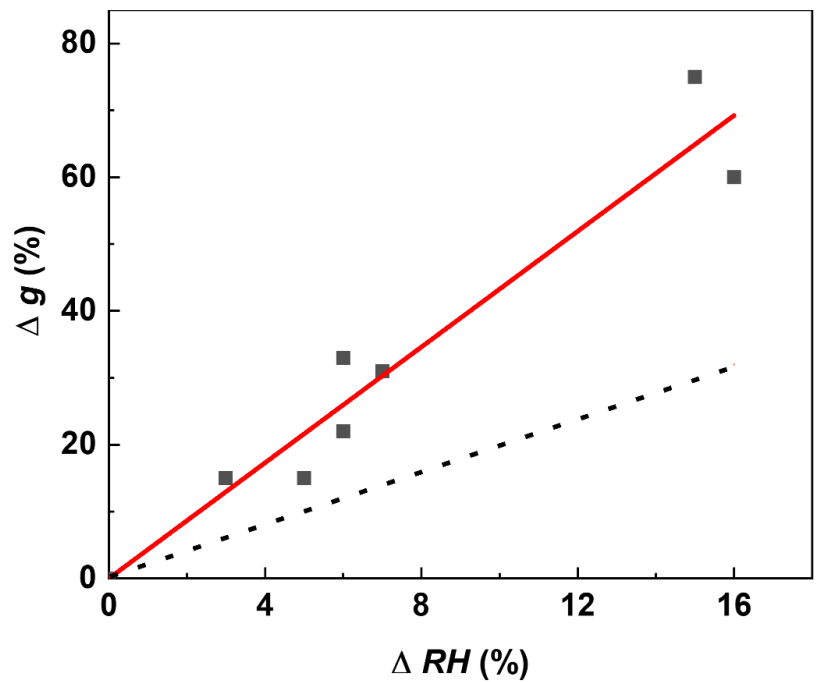

Figure 6. Variation of evaporation rate *g* with relative humidity *RH*, plotted as change of *g* versus change of *RH*. The dashed line with slope 1.0 would be expected for a linear relation between these quantities.

A final experiment was conducted in an OCA 25 droplet analyser to find out if there is any change of surface tension when Millipore water is exposed to a 450 mT field for 200 s. The result of 37 measurements was that the mean increase of the surface tension of water was barely significant, $0.19 \pm 0.21$ mNm$^{-1}$, a result that is consistent with previous work in a similar magnetic field [26], and not inconsistent with the increase of 1.31 mNm$^{-1}$ found in 10 T [27].

Summary of average water evaporation rates with and without magnetic field.

|  | Evaporation rate (kgm$^{-2}$h$^{-1}$) B = 500 mT | Evaporation rate (kgm$^{-2}$h$^{-1}$) B = 0 | *RH* (%) |
|---|---|---|---|
| Water | 0.0301 | 0.0274 | 72 |
| 6 M urea | 0.0321 | 0.0391 | 65 |

Initial and steady-state water evaporation rates and initial decay time in no field

|  | Initial evaporation rate (kgm$^{-2}$h$^{-1}$) | Decay time $t_0$ (minutes) | Steady evaporation rate (kgm$^{-2}$h$^{-1}$) | $T$ (°C) | $RH$ (%) |
|---|---|---|---|---|---|
| Water | 0.1360 | 9 | 0.0452 | 24 | 60 |
| 6 M urea | 0.0166 | 62 | 0.0310 | 24 | 60 |

## 4. Discussion

Ours is the first study where the effect of a uniform magnetic field on the evaporation rate of water or urea solution has been monitored over long periods simultaneously for an in-field sample and a no-field reference. Ambient laboratory temperature $T$ and relative humidity $RH$, which were monitored throughout, are the same for the in-field and no-field samples in each run. Evaporation rates increase with temperature and decrease sharply with relative humidity. There are also some uncontrolled fluctuations in the beakers, despite the long duration of our measurements, probably due to inevitable variations in surface airflow related to convection as water vapor evaporates and condenses at the water/air interface. Differences are observed between different measurement campaigns and even when the experimental setup was moved to a different place in the laboratory. Nevertheless, it is clear that there is an increase in the net evaporation rate of water of about 12% in the field, and a decrease in the net evaporation rate the urea solution that is at least as large. Table 4 with details of all the 16 and 60 hour runs is found in the Appendix. Furthermore, we have found significant differences in the transient and longer term evaporation of both water and urea. These are the data we seek to explain.

A commonly-used empirical formula that relates the net evaporation rate of water from a free surface in kgm$^{-2}$h$^{-1}$ to relative humidity $RH$ and the temperature-dependent dimensionless capacity $x_s$ of dry air to absorb water vapour is [28]

$$g = \Theta x_s (1 - RH) \qquad (2)$$

where $x_s$ is plotted in Supplementary Information. The pre-factor $\Theta$ is $(25 + 17v)$ where $v$ is the speed of the surface airflow in ms$^{-1}$. Units of $\Theta$ are kgm$^{-2}$h$^{-1}$. According to this formula, the anticipated evaporation rate from a free water surface in still air, $v = 0$, at $T = 22$ °C and $RH = 70\%$ is 0.150 kgm$^{-2}$h$^{-1}$. The evaporation rate increases by 10% whenever the temperature increases by 1.6 °C, the *RH* decreases by 5% or the surface air speed increases by 15 cms$^{-1}$.

The initial no-field evaporation rate of water from the surface of the partly-filled 100 mL beakers was 0.136 ± 0.045, close to that predicted by Eq. 2 with $\Theta = 25$, but it fell back to about a third of the initial rate value after about 15 minutes, as shown in Figure 6 and Table 3, Figure 9 gives evaporation rates based on final weight loss for 16 hour runs where the temperature of 20.8 – 23.0 °C and *RH* of 54 – 79% remained fairly constant. Humidity is an important variable, particularly since the data in Fig. 6 showed that a 5% change in relative humidity produces a 21 % change in evaporation rate, not the 10% change predicted by Eq. 2. The relation between evaporation rate and relative humidity is quadratic, not linear. The considerable scatter of the data in Fig. 7 can be ascribed to differences in humidity and temperature as well as uncontrolled local convection at the surface of the water between runs. The slope of the lines here are $\Theta = 5.57 \pm 0.24$ kgm$^{-2}$h$^{-1}$ in zero field and $\Theta = 5.01 \pm 0.27$ kgm$^{-2}$h$^{-1}$ in 500 mT.

The average increase of 11 - 12 % in the evaporation rate of water in the magnetic field cannot be attributed to change of temperature in the course of a run; the average temperature difference between the magnetic and reference water was 0.1 K, and in no case did it exceed 0.3 K in any run (Fig. S2). The mass of water vapour, 10g/kg present in the 0.182 m$^3$ Perspex enclosure is 1.82 g, which is the amount produced by the water in the two beakers in 21 h. By the end of a 16 h run, much of the water vapour in the enclosure has been generated from the beakers. There was no evident trend of increasing relative humidity in the enclosure during a run, because it is not hermetically sealed and it exchanges air with the ambient atmosphere in the laboratory .

The evaporation rates of water from filled circular vessels of different diameters including one similar to our beakers, were studied by Hisatake et al[29,30] in a fixed airflow. They found that rates in an airflow of 90 cms$^{-1}$ are roughly double the ones we find for our 50 g samples in partly filled beakers where the water vapour escaping from the surface is not blown away.

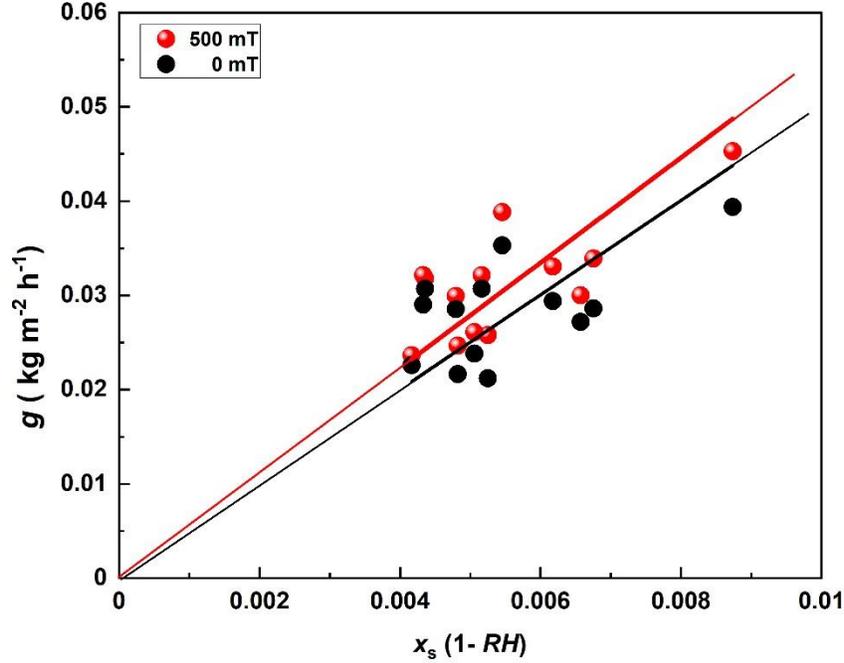

Figure 7. Evaporation rate of water from beakers containing 50 g of water measured at temperatures in the range 20.8 – 23.0 °C and *RH* of 56 – 79 % with or without a magnetic field. The scatter in the data is attributed to uncontrolled surface convection and a strong dependence on relative humidity.

The relatively small, steady evaporation rates we find in the beakers after an initial transient are attributed to the build-up of vapour that has a ortho/para isomeric ratio close to that of the vapour escaping from the liquid, rather than the equilibrium 3:1 ratio that was there in the ambient air at the beginning. The time taken to evaporate a volume of fresh water vapour to replace that originally in the beaker airspace is about 20 minutes, so in the extended runs vapour evaporating from the liquid surface with an isomeric ratio $f_L^o$: $f_L^p$ is expected to have the same isomeric ratio as the vapour in the beaker $f_V^o$: $f_V^p$, except at the beginning of the extended runs. Here p and o denote the para and ortho isomers and L and V denote freshly-evaporated vapour and the ambient vapour near the interface in the airspace. $f$ is the fraction of each isomer and $f^o + f^p = 1$ in any vapour.

We now propose an explanation of our data that treats the two nuclear isomers of water vapour as independent gasses. The properties of the nuclear isomers of diatomic hydrogen gas $H_2$ are well known[31]. Para-hydrogen with $I = 0$ is the more stable of the two. The $I = 1$ ground state of ortho-hydrogen has a quantum of rotational angular momentum, and lies 15.1 meV (175 K) higher in energy. It is possible to separate almost pure para-hydrogen gas at low temperature, although at room temperature the equilibrium ortho/para ratio is close to 3:1, reflecting the three-fold nuclear spin degeneracy of the $I = 1$ triplet. Interconversion between the two isomers is extremely slow in the gas phase, and non-equilibrium concentrations may persist in hydrogen for weeks.

Likewise there are two isomers of water, ortho-water with $I = 1$ and para-water with $I = 0$[32]. The ground state of ortho-water lies 35 K higher in energy. Relatively little is known about the difference in properties of the two isomers of liquid water. They were first identified by absorption onto charcoal surfaces, where unexpectedly long lifetimes of 26 and 55 minutes were reported[32,33], but these results were controversial [34]. Subsequently, Horke et al[35] produced pure beams of para- and ortho-water, and separated picolitre-scale volumes of each in a strong electric field. However, the separate molecular isomers will be much more stable in the gas phase, with little tendency to interconvert. The equilibrium 3:1 ratio will eventually be established by collisions after a period of many days, depending on vapour pressure. For our purposes, we will regard them as independent gasses, each with its own vapour pressure. It is thought that the ortho/para ratio in liquid water and the solutions we investigated may be far from the equilibrium value due to hydrogen bond formation[32]. The ratio has been investigated by four-wave mixing experiments[36][37], terahertz spectroscopy [38][39], nuclear magnetic resonance [37] and electrochemical impedance spectroscopy [40]. The ortho/para ratio can be enriched in water that has undergone ultrasonic cavitation relative to its value in liquid water[37] where the value is thought to be close to 1:1 [36]. There are important implications of the ortho/para ratio for hydrogen bonding and the structure of water, as well as biological[36], physiological and climatic consequences [41]. The para isomer, with no rotational angular momentum in the ground state, was reported to adhere preferentially to solid surfaces [35].

The saturation concentration of water vapour $x_s$ shown in Fig. S3, will be practically independent of the isomer involved but, when the two isomers behave as distinct gasses, the evaporation of each one is limited by its own partial pressure in the ambient air. The evaporation rate depends on the product of an escape probability related to the concentration of the isomer in the liquid, and a factor depending on the unsaturated vapour pressure of the escaping isomer in the ambient air in the beaker. If the ortho/para ratio for the molecules *escaping* from the water is $f_L$:(1-$f_L$) and $f_V$:(1-$f_V$) for the molecules in the vapour in the beaker, then we should consider a sum of the two independent fluxes of ortho and para vapour, which we do by introducing the factor γ in square brackets in equation 2.

$$g' = \Theta x_s (1 - RH)[f_L^0(1-f_v^0) + f_L^p(1-f_v^p)] \qquad (2)$$

We consider two limits. One is the case when the water is evaporating into normal 3:1 vapour from the atmosphere at the beginning of a run. The other is when the beaker is filled with vapour of the same isomeric composition as that released from the surface of the liquid itself. In the first case,

$$\gamma = 1 - f_L^0 f_v^0 - f_L^p f_v^p \qquad (3)$$

Writing this in terms of the ortho isomer, using $f_{L/v}^p = 1 - f_{L/v}^o$, and setting $f_v^o = 0.75$, we find the linear relation shown by the red line in Fig. 8.

$$\gamma = 0.75 - f_L^0 f_v^0 \qquad (4)$$

In the second case $f_L^0 = f_v^0$, and Eq. 3 reduces to

$$\gamma = 2 f_L^0 (1 - f_L^0), \qquad (5)$$

shown by the green parabola in the same figure. The evaporation rates in the two cases are proportional to γ, read off the figure for a given value of $f_L^0$. When the initial evaporation rate is greater than it is in the steady state, as in the case in pure water, the isomeric composition of the vapour released $f_L^0$ must lie in zone $A_1$ or zone $A_2$. When the converse is true, as in 6 M urea solution, $f_L^0$ must lie in zone B. To progress further with our model, and reproduce numerical ratios of the red and green evaporation rates, we have to bring condensation into the picture[42,43]. Equation (1) describes the *net* rate, evaporation less condensation. Equation (2) describes the two-gas evaporation. To model condensation, we will simply assume a constant value c independent of $f_L^0$ and add -c to the right hand side of Eq. 4 and Eq. 5. The value of *c* is indicated by the horizontal dashed line in the figure, which is the baseline from which we measure the net evaporation rate For water, we can reproduce the values in Table 3 for values of $1.05 < c < 0.45$ in zone $A_1$ and $0.1 < c < 0.2$ in zone $A_2$. For Urea, the limits are much narrower $0.40 < c < 0.45$. If we further require that c and the equilibrium evaporation rate is similar in the two cases, we can narrow down the ranges of $f_L^0$ to $0.39 \pm 0.01$ in water and $0.60 \pm 0.05$ in 6M urea.

The urea molecule is hydrogen bonded to five surrounding water molecules in the solvation structure of urea. One of the water molecules shares two hydrogen bonds with the urea. Concentrated solutions of urea (6 M urea is 36 % urea by weight) poses a constraint to rotational dynamics of the water molecules in the solution [44], which might influence the equilibrium fraction of ortho-water, and the nature of the escaping molecules.. Urea itself has a very low vapour pressure, and does not evaporate significantly at ambient temperature.

We emphasise that the ratios we measure are not necessarily the ratios present within the liquids. Water molecules are ejected in a process that involves the coordinated making and breaking of hydrogen bonds of at least three molecules at the interface to provide the 10 -12 $k_B T$ of energy needed for one molecule to break a hydrogen bond and escape [45]. The high energy of the escaping molecule, ~ 3000 K, may tend to thermalize the isomer ratio. It is unclear exactly what fraction of ortho water is present in the liquid, but in pure water it seems likely that $f_L^0 < 0.5$.

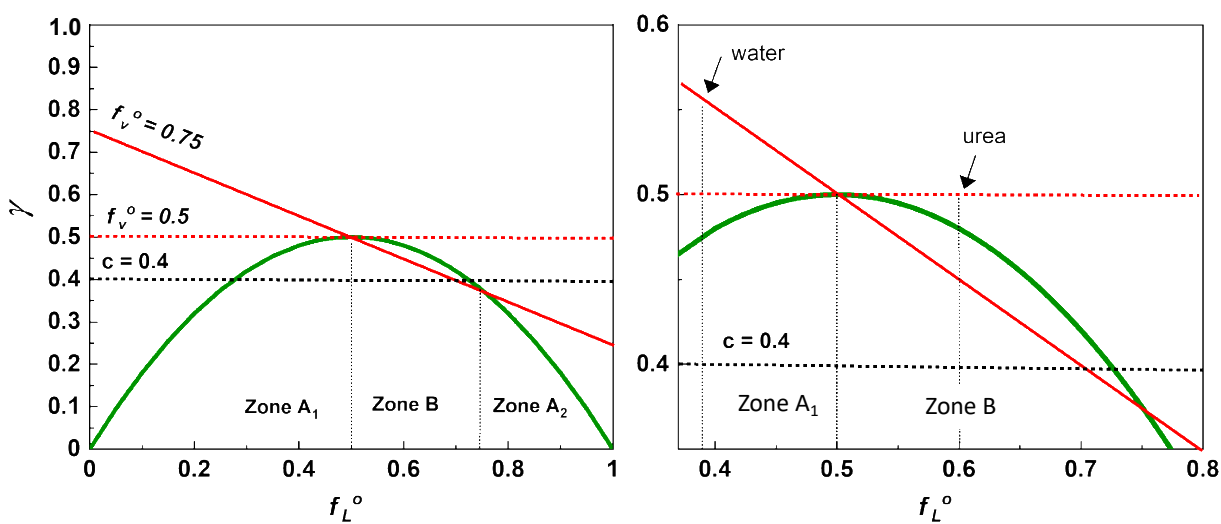

Figure 8: Normalized evaporation rates of water as a function of the ortho fraction $f_L^0$ present in the vapour escaping for the liquid for two cases. The red curve is for $f_V^0 - 0.75$, the natural 3:1 ratio in the atmosphere. The green curve is for $f_V^o = f_V^p$, when the evaporating liquid is surrounds by its own vapour. The isomeric ratios deduced from our data on pure water and 6M urea solution are marked (See text for details)

Finally, we consider two ways, illustrated in Fig. 9, whereby a magnetic field might modify $f_V^o$ at room temperature and thereby influence the evaporation rate. One possibility is via Larmor precession of the two hydrogen nuclei in a water molecule. Another is via Lorentz stress on the electric dipole moment of water. Protons precess at a frequency of 43 MHz/T, or 22 MHz in our 500 mT field. The small field gradient ~ 3 Tm$^{-1}$ near the edge of the beaker means that the precession frequencies for the two protons separated by 0.27 nm differ by about 1 part in $10^{10}$. The time taken for their precession to dephase by $\pi/2$ is 7 s, during which time the molecule will have travelled 4.4 km at an average speed of 630 ms$^{-1}$ and undergone N = $4.4 \times 10^{10}$ collisions if the mean free path is λ = 100 nm. In the 7s, the randomly-scattered molecule will have only drifted a distance of approximately √(N/3) λ = or 1.2 cm away from the liquid surface, while remaining in the beaker. Collisions do not influence the Larmor precession, which is independent of the orientation of the protons relative to the magnetic field. If the effect of the Larmor spin flips is to equalize the populations of the two isomers in the vapour, and the composition of the vapour $f_V$ is the same as that of the molecules escaping from the water $f_L$ in the absence of a field, the evaporation rate will always be enhanced by the field, because the green curve always lies below the dashed red line plotted $f_V^0 = f_V^0 = 0.5$.

Lorentz stress arises because the water molecule has a dipole moment p = $6.2 \times 10^{-30}$ Cm, The force on the moving electronic charge will tend to reduce the 105° bond angle. The torque on the molecule p$B$v is equal to the rate of change of angular momentum, d$l$/dt. A quantum of angular momentum is transferred to the water molecule in a time $\hbar$/pv$B$ = $5.4 \times 10^{-8}$ s. However, in that time the molecule undergoes on average 2400 collisions. Unlike Larmor precession, which is independent of the direction of travel of the protons and their orentation relative to the field, these contributions add statically and ~ 20 s will be required to add or subtract from the quantized angular momentum of the molecule, The tendency however will be to increase the ortho population, and in this case it is possible to diminish the evaporation rate in the field (Fig. 10), provided the composition lies in zone B, as is the case for the urea solution. The evapoation rate for water increases if it lies in zone A$_1$. The Larmor precession cannot explain the different signs of the field effects on evaporaton, the Lorentz torque could.

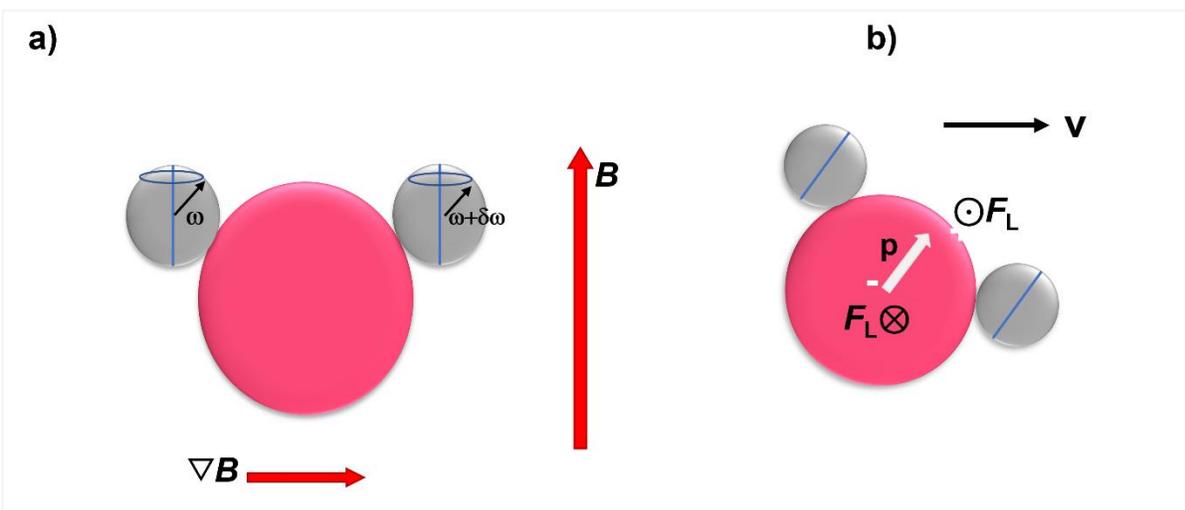

Figure 9. Modification of the angular momentum of a water molecule a) by dephasing of proton precession in a magnetic field gradient and b) by Lorentz torque due to motion of the charge dipole across the magnetic field

## 5. Conclusions

The rapid short-term decrease in evaporation rate of water and increase the evaporation rate of urea solution are attributed to the replacement of water vapour of ambient air in the beaker with an equilibrium 3:1 ortho:para ratio by a vapour whose isomeric composition increasingly resembles the that of the vapour escaping from the liquid surface. A phenomenological model of evaporation of ortho and para water vapour that treats the two as independent gasses allows the water and the urea solution to be allocated to different zones of isomeric composition. Narrowing down the possible regions of composition by including condensation and assuming the same value c ≈ 0.45 for both liquids leads to estimates of the ortho fraction $f_L^0$ in vapour freshly-evaporated from liquid of 0.39 ± 0.01 in water and 0.60 ± 0.05 in 6M urea solution. A slightly different, acceptable choice of *c* makes little difference to the numbers.

Our extensive studies in runs lasting up to 60 hours of evaporation of water exposed to a 500 mT magnetic field, with a no field control measured simultaneously side-by-side in the same environment show a consistent *increase* of approximately 12 % in evaporation rate in the magnetic field, under conditions where evaporation is not greatly enhanced by surface airflow. Fluctuations and scatter in the date are mostly ascribed to variations in ambient humidity, which has a strong influence on the evaporation rate. This confirms earlier reports [7,8,10,13,14,16] of an increase based on less rigorous methodology. For 6M urea the effect of the magnetic field is to produce a larger *decrease* of evaporation rate. To the best of our knowledge, this is a new finding. The different field effects on the evaporation are attributed to the ability of the field to alter the ortho:para ratio in the water vapour on the beaker. Of the two mechanisms we consider, nuclear Larmor precession will always increase the rate, regardless of isomeric composition, contrary to observation. Lorentz stress will always increase ortho:para ratio, and produce the opposite signs of the field effects we observe in water and urea.

Our model is phenomenological and simplified, but it makes clear predictions that can be tested in further experiments, for example, sufficiently strong magnetic field gradients should always enhance the evaporation rate. It is worthwhile examining terahertz spectra of rotational transitions on freshly evaporated water vapour in an effort to quantify the isomeric ratio in water vapour collected in confined spaces from different liquids, before and after magnetic treatment. Another experimental challenge is a controlled investigation on a timescale from minutes to hours of the remarkable claims of a magnetic memory in water and its possible relation to hydrogen bonding and the isomeric ratio in liquid water.

## Acknowledgements

Support from the European Commission from contract No 766007 for the 'Magnetism and Microfluidics' Marie Curie International Tranining Network and from Science Foundation Ireland contract 12/RC/2278 AMBER is acknowledged. We are grateful to Plamen Stamenov for helpful discussions and to Matthew Kavanagh for some of the data on urea.

# Appendix

Table 4 : Values of relative humidity, temperature and net evaporation rate of water with and without magnetic field

| Run number | Time (h) | Evaporation rate (kgm$^{-2}$h$^{-1}$) B = 500 mT | Evaporation rate (kgm$^{-2}$h$^{-1}$) B = 0 mT | Ratio | Relative Humidity (%) | Temperature (°C) |
|---|---|---|---|---|---|---|
| 1 | 16 | 0.0258 | 0.0212 | 1.216 | 72.4 | 21.7 |
| 2 | 16 | 0.0318 | 0.0307 | 1.035 | 78.0 | 22.4 |
| 3 | 16 | 0.0237 | 0.0226 | 1.047 | 78.6 | 22.1 |
| 4 | 16 | 0.0321 | 0.0290 | 1.108 | 78.4 | 22.5 |
| 5 | 16 | 0.0321 | 0.0307 | 1.047 | 74.6 | 22.8 |
| 6 | 16 | 0.0339 | 0.0286 | 1.185 | 63.6 | 21.3 |
| 7 | 16 | 0.0300 | 0.0272 | 1.104 | 63.9 | 21.0 |
| 8 | 16 | 0.0388 | 0.0353 | 1.100 | 69.6 | 20.8 |
| | | | | | | |
| 9 | 16 | 0.0254 | 0.0194 | 1.309 | 65.8 | 20.4 |
| 10 | 16 | 0.0540 | 0.0452 | 1.195 | 66.0 | 19.9 |
| | | | | | | |
| 11 | 16 | 0.0335 | 0.0321 | 1.044 | 68.4 | 21.5 |
| 12 | 16 | 0.0353 | 0.0325 | 1.087 | 67.3 | 21.2 |
| 13 | 16 | 0.0410 | 0.0378 | 1.084 | 71.1 | 23.1 |
| 14 | 16 | 0.0247 | 0.0219 | 1.129 | 77.1 | 22.2 |
| 15 | 16 | 0.0212 | 0.0191 | 1.111 | 80.7 | 22.4 |
| 16 | 16 | 0.0240 | 0.0230 | 1.046 | 81.0 | 22.9 |
| 17 | 16 | 0.0272 | 0.0247 | 1.100 | 78.1 | 22.4 |
| | | | | | | |
| 18 | 16 | 0.0222 | 0.0194 | 1.145 | 77.1 | 22.3 |
| 19 | 16 | 0.0247 | 0.0240 | 1.029 | 71.1 | 21.8 |
| 20 | 16 | 0.0293 | 0.0265 | 1.107 | 74.8 | 23.3 |
| 21 | 16 | 0.0240 | 0.0205 | 1.172 | 78.8 | 22.6 |
| 22 | 16 | 0.0222 | 0.0208 | 1.068 | 76.4 | 22.6 |
| 23 | 16 | 0.0318 | 0.0233 | 1.364 | 76.4 | 22.5 |
| | | | | | | |
| 24 | 60 | 0.0299 | 0.0285 | 1.050 | 76.0 | 22.5 |
| 25 | 60 | 0.0453 | 0.0394 | 1.151 | 56.0 | 22.4 |
| 26 | 60 | 0.0247 | 0.0217 | 1.139 | 76.0 | 22.6 |
| 27 | 60 | 0.0331 | 0.0294 | 1.125 | 66.9 | 21.4 |

| Run number | Time (h) | Evaporation rate (kgm⁻²h⁻¹) B = 500 mT | Evaporation rate (kgm⁻²h⁻¹) B = 0 mT | Ratio | Relative Humidity (%) | Temperature (°C) |
|---|---|---|---|---|---|---|
| 28 | 60 | 0.0261 | 0.0238 | 1.095 | 75.7 | 23.2 |
|  |  |  |  |  |  |  |
| 29 | 16 | 0.0543 | 0.0512 | 1.061 | 61.8 | 22.9 |
| 30 | 16 | 0.0483 | 0.0459 | 1.052 | 70.9 | 21.2 |
| 31 | 16 | 0.0391 | 0.0377 | 1.037 | 67.0 | 21.4 |
| 32 | 16 | 0.0370 | 0.0338 | 1.095 | 78.1 | 22.9 |
|  |  |  |  |  |  |  |
| 33 | 16 | 0.0406 | 0.0349 | 1.163 | 75.3 | 23.1 |
| 34 | 16 | 0.0451 | 0.0402 | 1.122 | 70.7 | 23.8 |
| 35 | 16 | 0.0353 | 0.0296 | 1.193 | 79.3 | 23.5 |
| 36 | 16 | 0.0427 | 0.0377 | 1.133 | 78.3 | 25.6 |
|  |  |  |  |  |  |  |
| **Average** |  | 0.0331 | 0.0297 | 1.118 | 72.8 | 22.3 |
| **Standard deviation** |  | 0.0088 | 0.0081 | 0.074 | 6.01 | 1.04 |

Table 5 : Values of relative humidity, temperature and net evaporation rate of 6M urea solution with and without magnetic field

| Run number | Time (h) | Evaporation rate (kgm⁻²h⁻¹) B = 500 mT | Evaporation rate (kgm⁻²h⁻¹) B = 0 mT | Ratio | Relative Humidity (%) | Temperature (°C) |
|---|---|---|---|---|---|---|
| 1 | 16 | 0.0314 | 0.0399 | 0.788 | 50.5 | 22.8 |
| 2 | 16 | 0.0307 | 0.0427 | 0.719 | 46.3 | 21.7 |
| 3 | 16 | 0.0392 | 0.0452 | 0.867 | 48.0 | 21.6 |
| 4 | 16 | 0.0307 | 0.0410 | 0.750 | 51.0 | 21.7 |
|  |  |  |  |  |  |  |
| **Average** |  | 0.0330 | 0.0422 | 0.781 | 49.0 | 22.0 |
| **Standard deviation** |  | 0.0041 | 0.0023 | 0.064 | 2.2 | 0.6 |

Table 6 : Values of relative humidity, temperature and net initial and steady evaporation rates of water and 6M urea solution

| Run | Time (h) | Liquid | Intial rate of evaporation (kgm$^{-2}$h$^{-1}$) | Steady rate of evaporation (kgm$^{-2}$h$^{-1}$) | Relative Humidity (%) | Temperature (°C) |
|---|---|---|---|---|---|---|
| 1 | 1 | water | 0.185 | 0.0621 | 54.5 | 25.0 |
| 2 | 1 | Water | 0.167 | 0.0876 | 54.5 | 25.0 |
| 3 | 1 | Water | 0.154 | 0.0734 | 55.0 | 25.0 |
| 4 | 1 | Water | 0.131 | 0.0702 | 55.0 | 25.0 |
| 5 | 1 | Water | 0.119 | 0.0720 | 57.5 | 24.5 |
| 6 | 1 | Water | 0.104 | 0.0705 | 57.5 | 24.5 |
| 7 | 1 | Water | 0.107 | 0.0620 | 62.0 | 24.5 |
| 8 | 1 | Water | 0.107 | 0.0523 | 62.0 | 24.5 |
| | | | | | | |
| 9 | 2 | water | 0.136 | 0.0452 | 60.0 | 24.0 |
| | | | | | | |
| **Average** | | | 0.134 | 0.0662 | 57.6 | 24.7 |
| **Standard deviation** | | | | | 3.1 | 0.4 |
| | | | | | | |
| 10 | 1 | 6M Urea | 0.0107 | 0.0427 | 73.0 | 24.2 |
| 11 | 1 | 6M Urea | 0.0141 | 0.0270 | 73.0 | 24.2 |
| 12 | 1 | 6M Urea | 0.0062 | 0.0257 | 68.5 | 23.5 |
| 13 | 1 | 6M Urea | 0.0085 | 0.0242 | 68.5 | 23.5 |
| | | | | | | |
| 14 | 2 | 6M Urea | 0.0166 | 0.0310 | 60.0 | 24.0 |
| | | | | | | |
| **Average** | | | 0.011 | 0.0301 | 68.6 | 23.9 |
| **Standard deviation** | | | 0.004 | 0.0070 | 5.3 | 0.4 |

# Supplementary Information

## Evaporation of water and urea solution in a magnetic field; the role of nuclear isomers

Sruthy Poulose, M. Venkatesan, Matthias Möbius and J.M.D. Coey

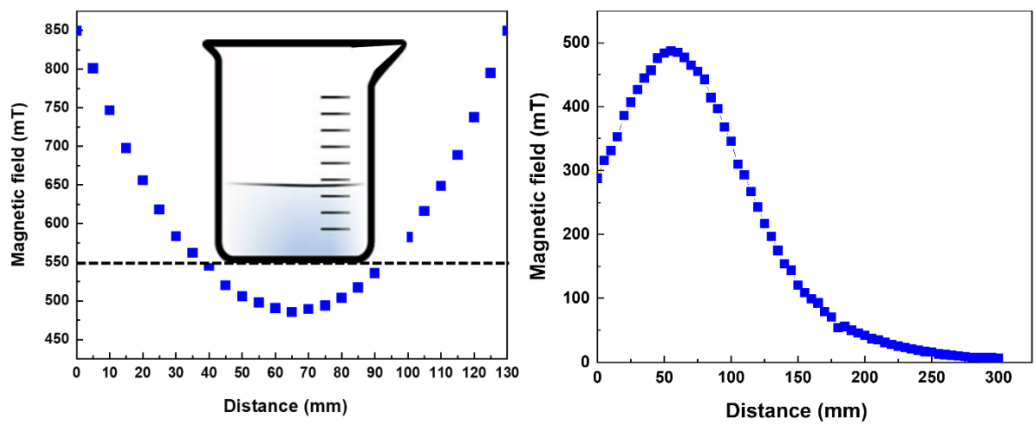

Figure S1 The in-plane field profile across the bore of the Halbach cylinder (left) and profile along its axis (right)

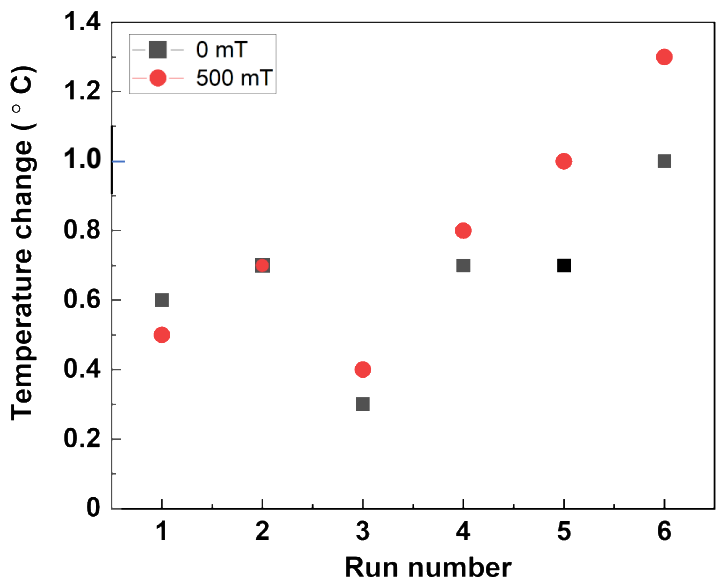

Figure S2 Change of temperature in the course of 16 h runs between in-field (red) and zero field (black) water. The average over all six runs in only 0.1 °C.

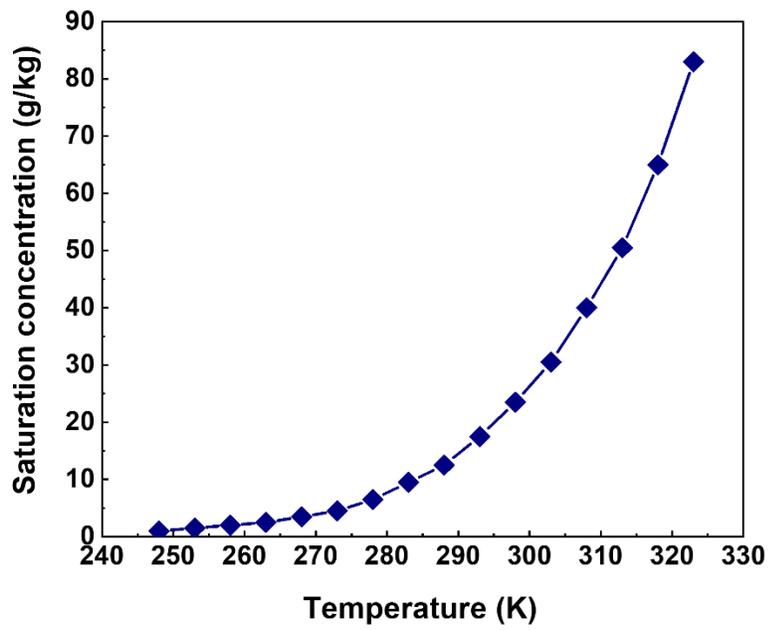

Figure S3 The saturation concentration of water vapour in air $x_s$ as a function of temperature [28].